# High-order fractal states in graphene superlattices

R. Krishna Kumar[1,2], A. Mishchenko[1,2], X. Chen[2], S. Pezzini[3], G. H. Auton[2], L. A. Ponomarenko[4], U. Zeitler[3], L. Eaves[1,5], V. I. Fal'ko[1,2*], A. K. Geim[1,2*]

[1]School of Physics & Astronomy, University of Manchester, Oxford Road, Manchester, M13 9PL, United Kingdom
[2]National Graphene Institute, University of Manchester, Oxford Road, Manchester, M13 9PL, United Kingdom
[3]High Field Magnet Laboratory (HFML), Radboud University, Toernooiveld 7, 6525 ED Nijmegen, Netherlands
[4]Department of Physics, University of Lancaster, Lancaster LA1 4YW, United Kingdom
[5]School of Physics and Astronomy, University of Nottingham NG7 2RD, United Kingdom

**Graphene superlattices were shown to exhibit high-temperature quantum oscillations due to periodic emergence of delocalized Bloch states in high magnetic fields such that unit fractions of the flux quantum pierce a superlattice unit cell. Under these conditions, semiclassical electron trajectories become straight again, similar to the case of zero magnetic field. Here, we report magnetotransport measurements that reveal second-, third-, and fourth-order magnetic Bloch states at high electron densities and temperatures above 100 K. The recurrence of these states creates a fractal pattern intimately related to the origin of Hofstadter butterflies. The hierarchy of the fractal states is determined by the width of magnetic minibands, in qualitative agreement with our band structure calculations.**

For electrons in solids, their ability to freely propagate through the crystal lattice (Fig. 1a) originates from the translational invariance of the Hamiltonian associated with a periodic lattice potential. According to Bloch's theorem[1], the electronic states are dispersed in energy and described by wave functions that are delocalized over the entire crystal lattice. This description generally breaks down in the presence of magnetic field ($B$), because electrons experience a Lorentz force and become localized on closed orbits[2-5] (Fig. 1b). However, for certain $B$ where the magnetic length is commensurable with the lattice periodicity (Fig. 1c), electrons recover delocalized wavefunctions[6,7] and behave as "magnetic" Bloch states that propagate along open trajectories as if they are effectively in zero magnetic field ($B_{eff} = 0$). Mathematically, this occurs for all rational fractions of magnetic flux $\phi = SB = \phi_0 p/q$ where $S$ is the area of crystal's unit cell, $\phi_0$ is the flux quantum, and $p$ and $q$ are integer numbers[6,7]. Physically, this is a consequence of the Aharonov-Bohm effect so that an electron passing across $q$ unit cells acquires a phase shift in multiples of $2\pi$, which restores the translational periodicity in high $B$. The recurrence of propagating Bloch states is expected to cause fractal, self-similar behavior in the magnetotransport properties of crystalline solids[6-8].

In graphene/hexagonal boron-nitride (hBN) superlattices[9-12], the periodic structure created by recurrent magnetic Bloch states was previously observed in low temperature experiments[13-18], with most of the attention being paid to the detection of Landau gaps in the Hofstadter butterfly spectrum[8,19]. In particular, third generation Dirac points[13], the anomalous quantum Hall effect[14] and replica quantum Hall ferromagnetism[16] were observed in the magnetic Bloch states that resided at unit fractions of $\phi/\phi_0 = 1/q$. However, this periodicity does not constitute the complete self-similarity inherent to magnetic Bloch states[6,7] and Hofstadter butterflies[8,19], where the spectra should replicate themselves at increasingly smaller scales of $B$. This fractal structure can appear only due to high-order magnetic Bloch states with $p > 1$ and has awaited experimental confirmation.

In this Letter we probe the electronic spectrum of graphene/hBN superlattices and report the full hierarchy of magnetic Bloch states up to the 4$^{th}$ order ($\phi/\phi_0 = 1/q$, $2/q$, $3/q$ and $4/q$) using an alternative approach that is based on the use of transport measurements at high temperatures ($T = 100 - 200$ K). In this regime, magnetotransport still reflects the characteristic properties of magnetic minibands but is not obscured by overlaying Shubnikov-de Haas (SdH) oscillations[20]. Our recent work[21] showed that magnetic Bloch states become most prominent above 100 K, resulting in robust quantum oscillations in magnetoconductivity ($\sigma_{xx}$). These so-called Brown-Zak (BZ) oscillations originate from the repetitive formation of Bloch states at magnetic fields which follow the sequence $\phi/\phi_0 = 1/q$. Upon increasing $B$, electron trajectories are modulated between closed and open orbits, which cause the conductivity to oscillate. In the experiments below, we extend the parameter space to high carrier densities ($n$) and $B$ up to 30 T, which allows the observation of a fractal pattern in $\sigma_{xx}$ originating from high-order magnetic Bloch states ($p > 1$).

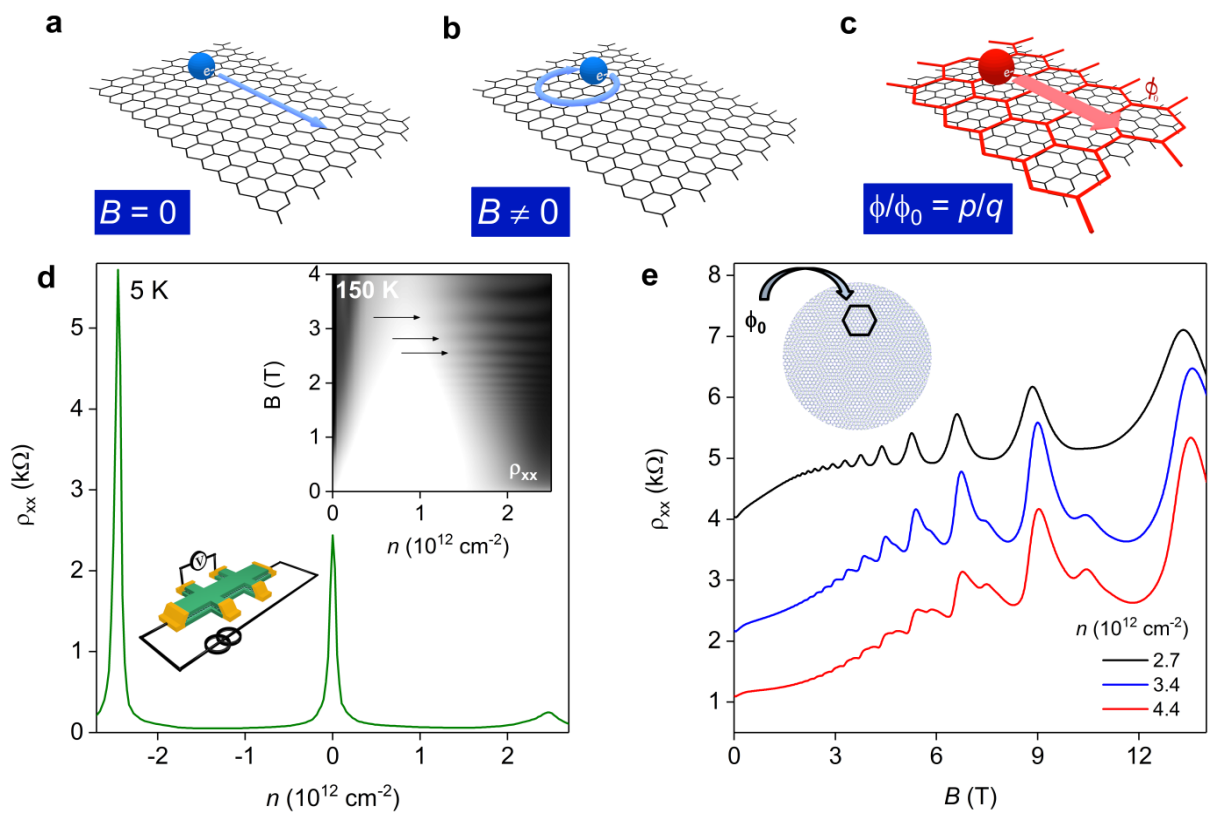

**Figure 1| Brown-Zak oscillations in graphene/hBN superlattices. a-c**, Schematic illustration of electron trajectories for different magnetic fields. The blue spheres (**a, b**) are electrons propagating along trajectories indicated in blue. Graphene's crystal lattice is shown as gray hexagons. The red sphere (**c**) represents a quasiparticle (magnetic Bloch state) propagating along a straight trajectory on a lattice of supercells (red hexagons), as if in zero effective magnetic field. Note that in the presence of an electric field trajectories in (c) may become curved and develop into chiral edge states because of nontrivial topology of the magnetic minibands[33-35,39]. **d,** $\rho_{xx}(n)$ for a graphene superlattice with the period of ∼ 14 nm. Inset: Map $\rho_{xx}(n,B)$ for electron doping. Logarithmic gray scale: white 80 Ω; black 1,200 Ω. Lower inset: Device and measurement schematics. **e,** $\rho_{xx}(B)$ at 150 K for three carrier densities above $n_0$. Inset: Illustration of a graphene/hBN moiré superlattice. For $n = 2.7 \times 10^{12}$ cm$^{-2}$, the oscillations exhibit a single periodicity, which corresponds to one $\phi_0$ piercing the moiré unit cell (outlined by the black hexagon).

The studied devices were fabricated using the standard approach for making encapsulated graphene/hBN heterostructures[22] (for details, see Supplementary Section 1). During their assembly, a rotating stage was employed to accurately align graphene's crystallographic axes with the hBN substrate. The alignment resulted in a moiré superlattice[9-11,23] with a period of ≈ 14 nm due to a slight (~1.8%) mismatch between the graphene and hBN crystal lattices. This step is crucial to observe the physics described above, as it ensures that the regime with $\phi/\phi_0 = 1$ can be reached for $B$ below 30 T. Note that for pristine graphene (without a superlattice potential) the above condition would be met only at ~ 10,000 T. A second hBN crystal was placed on top of the graphene to encapsulate it, ensuring high electronic quality[24]. The top hBN was intentionally misaligned to avoid a competing moiré potential acting on graphene charge carriers. After the assembly, electron beam lithography and standard microfabrication processing were employed to etch the heterostructure into multiterminal Hall bar devices with quasi-one-dimensional contacts[25,26] (inset of Fig. 1d). We studied five superlattice devices and found the features described below in all of them but they were strongest in those with the highest electronic quality and largest moiré period.

Figure 1d shows an example of the measured resistivity ($\rho_{xx}$) in zero magnetic field as a function of $n$ for one of our superlattice devices. Three peaks in $\rho_{xx}$ are observed at $n = 0$ and $n = \pm 2.5 \times 10^{12}$ cm$^{-2}$. The latter two peaks provide an unambiguous indication of the superlattice reconstruction of graphene's spectrum[11,27,28] and are referred to as secondary Dirac points (DPs). Notably, the secondary DP is considerably sharper for hole doping (negative $n$) than electron doping (positive $n$), in agreement with the previous work[13-16] and with calculations that have demonstrated graphene's band structure is stronger modified for holes[27]. Rather surprisingly, BZ oscillations are found to be more pronounced for electrons in the conduction band[21], especially at high $T$. The inset of Fig. 1d plots a map of magnetoresistance $\rho_{xx}(n, B)$ for electron doping at 150 K. At this $T$, SdH oscillations and the corresponding Landau fans[13-16] are completely suppressed because of thermal smearing. Instead, we find a set of horizontal streaks across the map (highlighted by arrows), which signifies magneto-oscillations that are independent of $n$. These are Brown-Zak oscillations.

Figure 1e plots $\rho_{xx}(B)$ for several $n$ beyond the electron secondary DP. For $n = 2.7 \times 10^{12}$ cm$^{-2}$ (black curve), we find that $\rho_{xx}$ oscillates in $1/B$ with a single periodicity of $\phi_0/S$ (schematic of Fig. 1e). As we increase $n$ further, the oscillations start developing some extra periodicity. In particular, we find additional features appearing between the maxima. As shown below, these oscillations originate from the formation of magnetic Bloch states at $\phi/\phi_0 = 2/q$. Because the extra features become stronger at higher $n$ and higher $B$, whereas standard doping by electrostatic gating is limited to ~ $5 \times 10^{12}$ cm$^{-2}$, we employed fields up to 30 T and, at the same time, used optically-induced doping, a peculiar property of graphene/hBN heterostructures[29] (Supplementary Section 2). These techniques allowed us to reach $n$ as high as ~ $3n_0$ and perform measurement for $\phi/\phi_0 \geq 1$.

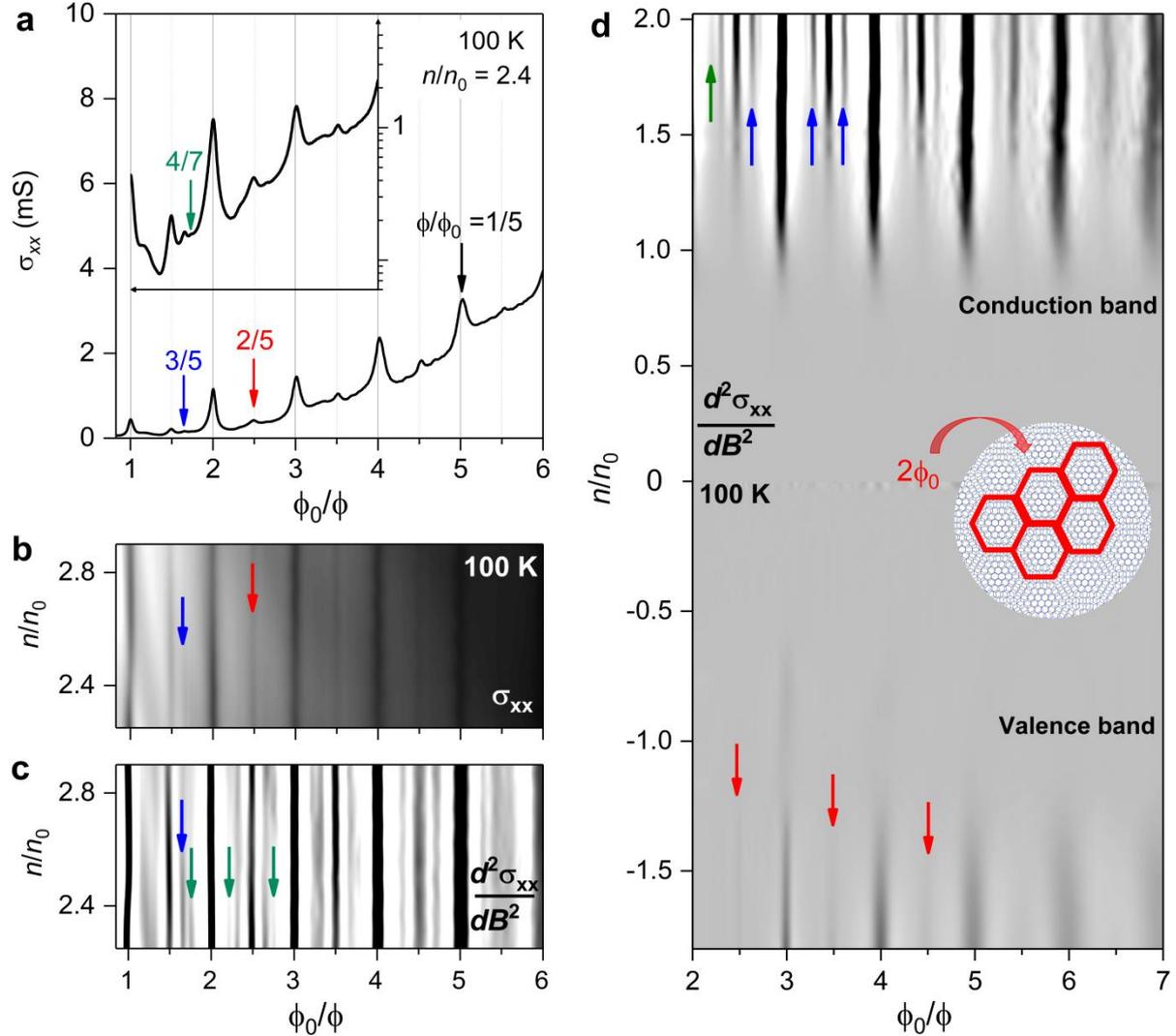

**Figure 2| High-order magnetic Bloch states. a,** $\sigma_{xx}$ as a function of $B$ expressed in units, $\phi_0/\phi$. Inset: Same data on a logarithmic scale to emphasize weak features; the same horizontal axes. **b,** $\sigma_{xx}(\phi_0/\phi,n)$ for high electron doping in $B$ up to 30 T ($\phi > \phi_0$) where fractal features are most visible. Logarithmic gray scale: white 0.05 mS; black 5 mS. **c,** Second derivative $d^2\sigma_{xx}/dB^2$ of the data in **b**. Gray scale: white 0 mS/T$^2$; black -0.05 mS/T$^2$. **d,** Same as in **c** but using another data set obtained in $B$ up to 15 T. The black, red, blue and green arrows mark fractions with $p$ = 1, 2, 3 and 4, respectively. The schematic in **d** illustrates the fractal state $\phi/\phi_0$ =2/5, which involves an extended unit cell of 5 original moiré unit cells that share two flux quanta between them.

It is instructive to analyze the BZ oscillations in terms of longitudinal conductivity $\sigma_{xx}(B) = \rho_{xx}/(\rho_{xx}^2 + \rho_{xy}^2)$ where $\rho_{xy}$ is the Hall resistivity. This is because $\sigma_{xx}$ exhibits local maxima at those $B$ where magnetic Bloch states emerge ($\phi/\phi_0 = p/q$) and quasiparticle trajectories in the superlattice potential become effectively straight again, mimicking transport at zero field[30-32] (Fig. 1c). In addition, the use of the dissipative conductivity $\sigma_{xx}$ simplifies our analysis by avoiding a nondissipative (Hall) contribution caused by topological properties of the magnetic minibands that can have non-zero Chern numbers[33-36]. Fig. 2a plots $\sigma_{xx}$ as a function of $\phi_0/\phi$, which emphasizes the $1/B$ periodicity of BZ oscillations. Here, we find local maxima in $\sigma_{xx}$ located at unit fractions of $\phi/\phi_0 = 1/q$, in agreement with the previous report[21]. In addition, several other maxima become clearly visible at fractions of $\phi/\phi_0 = 2/q$ and $3/q$. For high $B$, maxima at $\phi/\phi_0 = 4/q$ can also be discerned (inset of Fig. 2a, Fig. 2c).

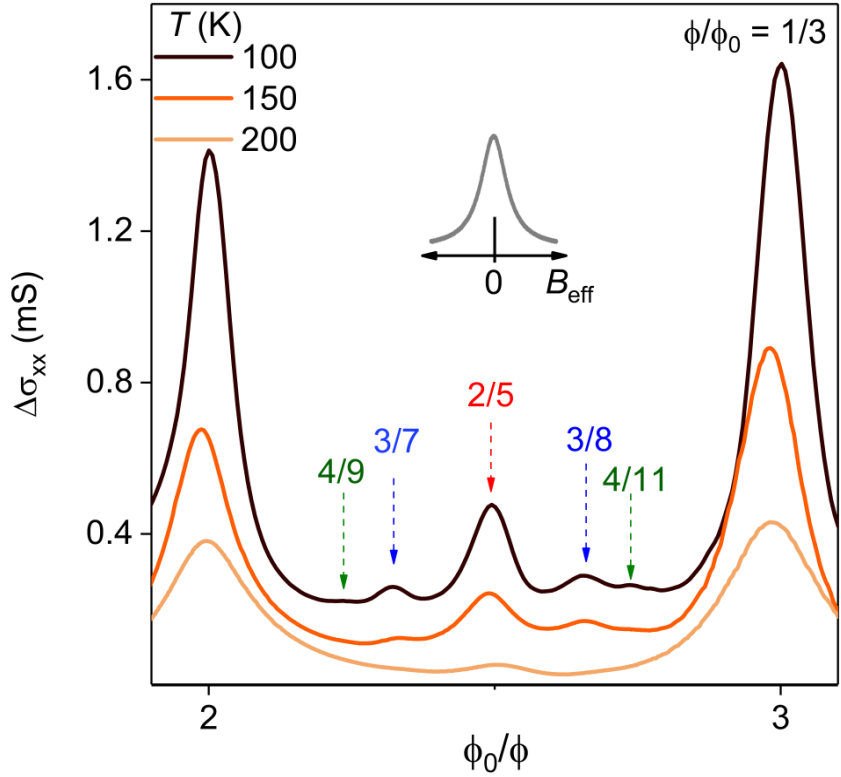

**Figure 3 | Temperature dependence of high-order states.** Longitudinal conductivity after subtracting a smooth background, $\Delta\sigma_{xx}(\phi_0/\phi)$. Three different $T$ for $n = 2.2\ n_0$. The arrows indicate fields where the fractal magnetic Bloch states with $p = 2$, 3 and 4 are expected. The curves are shifted vertically for clarity. Inset: Standard behavior of $\sigma_{xx}(B)$ for metallic systems near zero magnetic field, either applied ($B$) or effective ($B_{eff}$). The shape is described by the classical expression $\sigma_{xx} \propto 1/[1 + (\mu B)^2]$ where $\mu$ is the charge carrier mobility[31,32].

To better visualize the additional maxima with $p > 1$, we exploit one of the defining features of BZ oscillations, namely that their frequency is independent of $n$. Fig. 2b plots a map of $\sigma_{xx}(n, \phi_0/\phi)$, which reveals a set of dark vertical streaks. They can be seen more clearly if we plot the second derivative of $\sigma_{xx}$ with respect to $B$ (Fig. 2c). The differentiation procedure effectively removes the smooth background and highlights the extra features by sharpening local maxima. The vertical streaks indicate the extra features appearing at the same $B$ for all $n$. Even the maxima for $p = 4$ becomes clearly distinguishable as faint gray features independent of $n$ (green arrows in Fig. 2c). The observed behavior signifies that the additional maxima are caused by high-order magnetic Bloch states. Note that several $p = 2$ maxima can also be observed for hole doping (Fig. 2d), despite the poor visibility of BZ oscillations in graphene's valence band[21].

For certain ranges of $n$ and $B$, we were able to identify all the magnetic Bloch states up to a 4$^\text{th}$ order which can occur within the field interval $1/(q+1) < \phi/\phi_0 < 1/q$. This is illustrated in Fig. 3 for $q = 2$. At 100 K (black curve), the hierarchy of magnetic Bloch states creates a fractal pattern in the magnetoconductivity, that is, the behavior of $\sigma_{xx}$ close to zero applied magnetic field (inset in Fig. 3) is replicated multiple times at increasingly smaller scales of $B$. As $T$ increases, the fractions with large $p$ become smeared and only those with $p = 1$ remain (Fig. 3).

To understand the observed hierarchy of states, let us first recall how the energy spectrum of graphene superlattices looks in quantizing magnetic fields[13,37-39]. Fig. 4a plots the computed density of states in the conduction band as a function of energy and $B$, producing an image often referred to as the Hofstadter butterfly[8] (Supplementary Section 3). In general, the spectrum is dominated by localized states which are caused by Landau quantization. They occur at irrational values of $\phi/\phi_0$ and appear as numerous discontinuous regions marked by black dots. However, at rational $\phi/\phi_0 = p/q$ the spectrum becomes continuous due to the emergence of magnetic Bloch states[6,7]. These are represented by solid vertical lines in Fig. 4a and appear each time when $p$ flux quanta pierce a so-called supercell that has an area $q$ times larger than the moiré unit. For example, a magnetic Bloch state at $\phi/\phi_0 = 2/5$ (red line in Fig. 4a) arises if two flux quanta penetrate through a supercell that is 5 times larger than the moiré unit cell (inset of Fig. 2d). We also note that, for $n < n_0$, the superlattice spectrum is gapped over a wide range of $B$. This behavior is specific to Dirac electrons[37-39] and explains why BZ oscillations and high-order maxima in $\sigma_{xx}$ are absent in experiments performed at low $n$ (Fig. 2d).

The visibility of a particular magnetic Bloch state is determined by both the number of flux quanta ($p$) and the number of unit cells ($q$) associated with the state. As $q$ increases, its visibility is expected to decrease because the Bloch state involves an increasingly larger supercell that might not be fully traversable by electrons because of their limited mean free path. This explains why first-order states that follow the sequence $\phi/\phi_0 = 1/q$ (that is, BZ oscillations) tend to be more prominent (Fig. 1d,e). Note that this tendency does not always hold (see Fig. 2a for $\phi/\phi_0$ approaching unity and the discussion below). As for the $p$ dependence, its details are more subtle. When the supercell size is fixed by $q$, the maxima in $\sigma_{xx}$ still become progressively smaller with increasing $p$. This is evident from the sequence $\phi/\phi_0 = p/5$ shown in Fig. 2a (highlighted by the colored arrows). To understand the $p$ dependence, we recall[30-32] that transport of Bloch electrons also depends on their group velocity ($v$) as $\sigma_{xx} \propto v^2\tau$ where $\tau$ is the scattering time. Assuming that $\tau$ is independent of $B$ (ref. 32), maxima in $\sigma_{xx}$ should be determined by the value of $v$, which reflects the width of the energy bands of the corresponding states. Broad bands imply higher electron velocities than flat bands[31]. With this in mind, Fig. 4b shows the energy dispersion for the magnetic Bloch states at $\phi/\phi_0 = 1/5, 2/5$ and $3/5$ (same $q$ but different $p$). In general, the minibands contain many closely spaced subbands that either overlap or are separated by small gaps. As $p$ increases, the minibands become flatter and the gaps separating them more pronounced. This suggests that $v$ is smaller for magnetic Bloch states with larger $p$.

For further analysis, we calculated the group velocity for different magnetic Bloch states and, to account for the relatively high $T$ of our measurements, took an average over the interval of thermal smearing (Supplementary Section 4). Fig. 4c plots the resulting mean square velocity $<v^2>$ as a function of rational $\phi/\phi_0$ at a fixed $n$ for $p = 1, 2$ and $3$ and various $q$. Clearly, the states with larger $p$ have a systematically smaller $v$. For example, the average velocity for $\phi/\phi_0 = 2/5$ is ten times smaller than that for $\phi/\phi_0 = 1/5$. This proves that magnetic Bloch states with flatter bands exhibit lower average $v$ and, therefore, lower $\sigma_{xx}$. The latter makes it harder to resolve the states experimentally and explains the observed $p$ dependence of the local maxima in Figs. 2-3. We note that the calculated $<v^2>$ shows the same trend with $p$ for hole doping in the valence band (Supplementary Section 5). Furthermore, at small $q$, local maxima in $\sigma_{xx}$ become dependent not only on the supercell size (defined by $q$) but also on details of the mini-band structure, that is, on $<v^2>$. For example, the non-monotonic dependence of the first-order peaks in Fig. 2a ($\phi/\phi_0 = 1/q$) can be attributed to the interplay between the supercell size and $<v^2>$. Although magnetic Bloch states are formed more easily at small $q$ (high $B$), their average speed becomes significantly smaller (Fig. 4c), which reduces the local maxima in $\sigma_{xx}$.

Finally, we consider the $n$ dependence of high-order fractal states. The inset of Fig. 4c plots $<v^2>$ as a function of $n/n_0$ for the minibands shown in Fig. 4b ($\phi/\phi_0 = p/5$). For all the fractions, $<v^2>$ increases with $n$, in agreement with our experiment that shows more prominent fractal features at higher doping. The origin of higher $<v^2>$ in this case stems from the fact that the minibands become closely spaced at higher energies (Fig. 4a-b). Accordingly, the Fermi level becomes smeared over an increasing number of minibands, which in turn increases $\sigma_{xx}$ and, therefore, the visibility of magnetic Bloch states at high $n$. Note that some of the minibands are likely to have non-zero Chern numbers that can result in non-zero Hall conductivity in zero $B_{eff}$ and, at low $T$, in chiral edge states[33-34]. However, the topological properties should not affect the discussed dissipative $\sigma_{xx}$ in the linear response to the current-driving electric field.

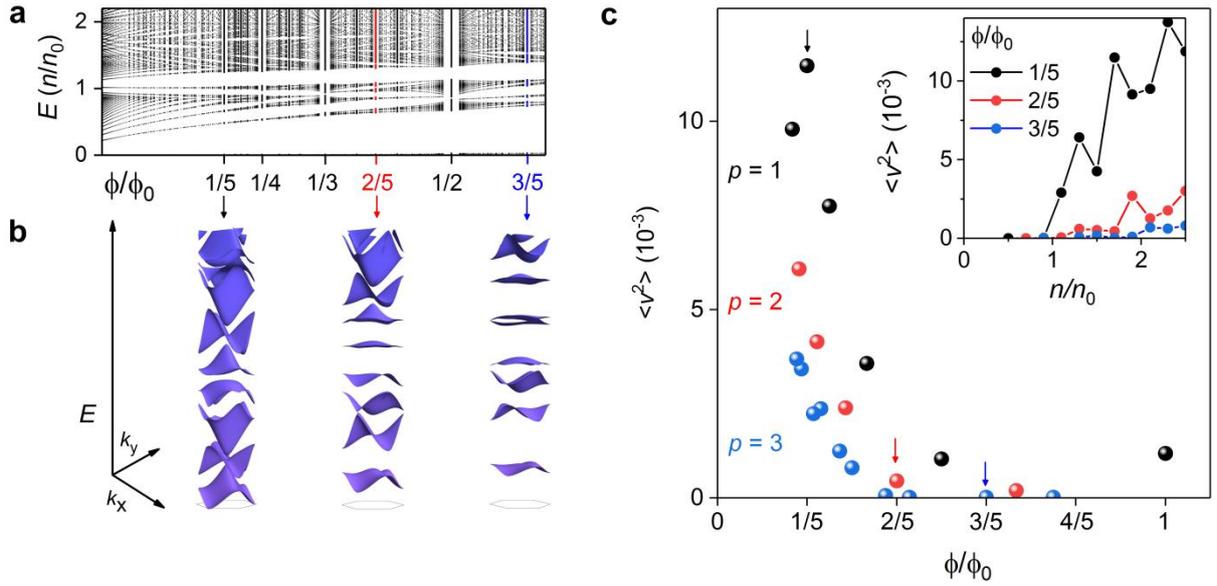

**Figure 4| Computed hierarchy of magnetic Bloch states**. **a,** A part of the Hofstadter butterfly for electrons in graphene/hBN superlattices[21,39]. The black and white regions signify allowed states and energy gaps, respectively. The red and blue vertical lines highlight the Bloch states at $\phi/\phi_0$ = 2/5 and 3/5, respectively. Note that the white vertical stripes around $\phi/\phi_0$ = $p/q$ indicate regions omitted in the calculations for technical reasons (too dense spectrum) [21,39]. **b,** Energy dispersions for the Bloch states with $\phi/\phi_0$ = 1/5, 2/5 and 3/5 over an energy interval from 0.2 to 0.3 eV, which approximately reflects the doping level in our measurements. **c,** Average group velocity $<v^2>$ for $n/n_0$ = 1.7. The values are normalized by graphene's Fermi velocity. Fractal states with different $p$ are color-coded. The black, red and blue arrows indicate the 1/5, 2/5 and 3/5 states shown in **b**. Inset: $<v^2>$ as a function of $n/n_0$ for those three states.

To conclude, in addition to Brown-Zak oscillations that are periodic in $1/B$ and correspond to $\phi = \phi_0/q$, magnetotransport in graphene superlattices exhibits a fractal pattern due to high-order magnetic Bloch states that are in principle expected for all rational $\phi/\phi_0 = p/q$ and are clearly observed in our experiments for $p$ = 2, 3 and 4. These high-order states require sufficiently high electron doping to become visible experimentally, and their hierarchy is associated with increasingly flatter Bloch minibands at higher $p$. Further work is required to understand the effect of topology and non-zero Chern numbers associated with magnetic minibands[39] on transport properties of graphene superlattices.

**Acknowledgements**

This work was supported by Engineering and Physical Sciences Research Council (EPSRC), the European Graphene Flagship, the Royal Society, the Lloyd's Register Foundation, the European Research Council (ERC) and the Netherlands Organisation for Scientific Research. A.M. acknowledges the support from the EPSRC Early Career Fellowship EP/N007131/1. R.K.K was supported by the EPSRC Doctoral Prize Fellowship (EP/N509565/1). U.Z & S.P acknowledge the support of HFML-RU/FOM, member of the European Magnetic Field Laboratory (EMFL). G.H.A acknowledges support from the EPSRC under the grant number EP/M507969/1. L.A.P acknowledges the Royal Society under the grant code UF120297. V.I.F was supported by the EPSRC (EP/N010345), the ERC Synergy Grant Hetero 2D (ERC-2012-SyG) & the European Graphene Flagship (H2020-SGA-FET-GRAPHENE-2017). R.K.K, G.H.A & V.F acknowledge support from the Graphene NOWNANO Doctoral Training Centre.

# Supplementary Information

**S1. Device fabrication**

Graphene/hexagonal boron-nitride (hBN) heterostructures were assembled using the dry peel technique[1,2]. To this end, graphite and bulk hBN crystals were first mechanically exfoliated onto an oxidized Si wafer. Monolayer graphene and thin (~30 nm) hBN flakes were then identified by optical microscopy. The flakes were assembled using a polymer membrane attached to the tip of a micromanipulator, which was used as a 'stamp' to pick up and place down the selected crystals. During their assembly, a rotating stage was employed to align crystallographic axes of graphene and the bottom hBN crystal. Because the bulk crystals cleave preferentially along their crystallographic axes, the edges are usually clear and straight, enabling alignment with accuracy of about $0.5^\circ$. The resulting alignment produces a moiré potential with a period of about 14 nm[3].

The heterostructures were then made into multiterminal devices such as shown in Fig. S1 using electron beam lithography and plasma etching. As the first step, a PMMA mask was fabricated to define long contact regions leading to the heterostructure. Reactive ion etching was then employed to mill through the mask, which produced trenches in the graphene/hBN heterostructure. The same PMMA mask was subsequently used to deposit metal leads into the trenches (3 nm Cr/80 nm Au) which formed quasi-one dimensional contacts to graphene's edges[4,5]. This sequence of steps mitigates the need for graphene to be in contact with any polymer, preserving its high electronic quality. The PMMA mask was removed and a second round of lithography was carried out to define the mesa. Figure S1 shows an example of one of our Hall bar devices. The width of our samples ranged between 1 and 3 μm, and the distance between nearest voltage contacts was at least one width.

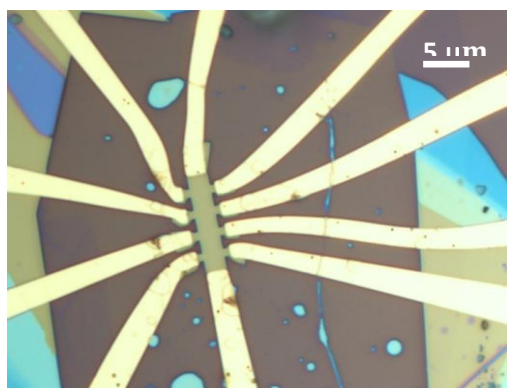

**Figure S1| Graphene/hBN devices.** An optical image of one of our superlattice devices. The width of the Hall bar is 3 μm.

## S2. Light doping of graphene/hBN heterostructures

The first order magnetic Bloch states, which are responsible for so-called Brown Zak oscillations, are robust over a wide range of temperatures ($T$) and carrier density ($n$). In contrast, even for $n$ close to the secondary Dirac points (DPs), there are no signs of high-order states ($p$ = 2, 3 etc.). To unveil those additional fractal states, we had to dope graphene to much higher $n$, close to and even beyond $n = 2n_0$ (Fig. 2 of the main text), where $n_0$ is the carrier density at which secondary DPs occur. To achieve such high n-type doping, we used light illumination as previously reported[6]. In brief, electron donor-like impurity states in hBN are excited by illumination. This leads to positively charged defects that act to dope the graphene sheet negatively. This may happen in both top and bottom hBN layers of our encapsulated devices. Because the defects in hBN are spatially isolated from the graphene channel, the doping has relatively little influence on mobility of charge carriers in the devices. The light-doping effect is sufficiently strong such that we were able to n-dope graphene by $n = 2.5 \times 10^{12}$ cm$^{-2}$ simply using an incandescent light source. Fig. S2 shows the electric-field behavior of our graphene/hBN superlattice devices before and after illumination. Before illumination, the main Dirac point (DP) is found around the applied gate voltage $V_G$ = - 1 V, whereas the secondary DPs are at $\pm$ 15 V. After illumination, the main DP shifted to -20 V, the electron secondary DP to -4 V and the hole secondary DP could not be reached by the electric field doping. Despite this large photo-induced doping, the electronic quality remained high with relatively small degradation in mobility. This is evident from the inset of Fig. S2 which plots the resistivity $\rho_{xx}$ as a function of $V_G - V_{DP}$. Aside from the slight broadening of the main Dirac point after illumination, the curves closely follow each other, in agreement with the previous work[6].

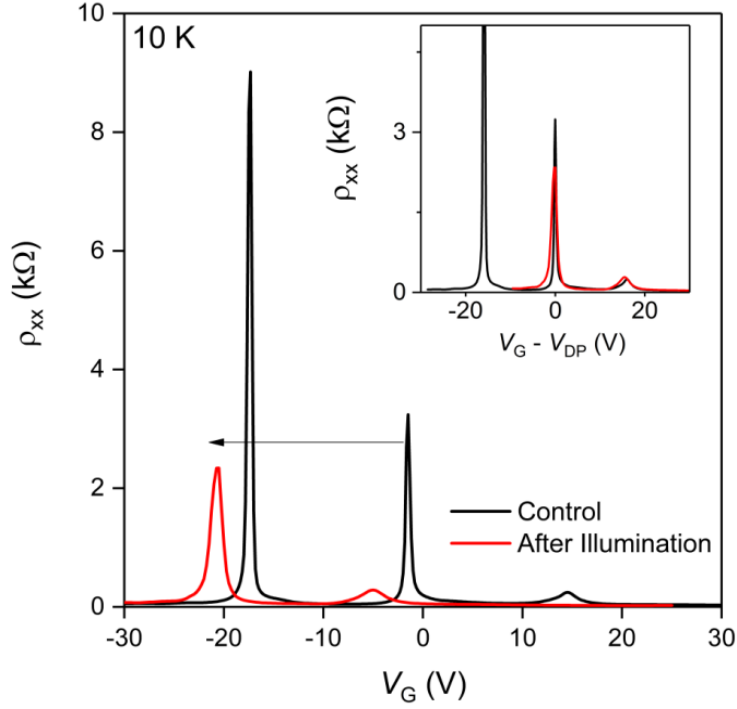

**Figure S2| Light doping of graphene/hBN heterostructures.** Longitudinal resistivity ρ_xx as a function of gate voltage $V_G$ for one of our graphene devices presented in the main text. Two curves are plotted; before (black curve) and after (red) illumination with an incandescent light source. The black arrow traces the shift in position of the main Dirac point after illumination. Inset; The same data plotted as a function of $V_G - V_{DP}$, where $V_{DP}$ is the position of the main Dirac point.

## S3. Magnetic mini-bands

Here we describe the model used to calculate the Hofstadter butterfly spectrum and magnetic mini-band structures shown in Figs. 4a-b of the main text. To this end, we consider the electronic spectrum of pristine graphene modified by an underlying moiré potential which is produced by alignment with the hBN substrate. The moiré potential is described by a hexagonal Bravais lattice $n_1 \vec{a}_1 + n_2 \vec{a}_2$ with a period of the superlattice $a_1 = a_2$. The computed spectrum was obtained using a phenomenological model developed in Ref. 7, which is based on the Hamiltonian.

$$\hat{H} = v_F \vec{p} \cdot \vec{\sigma} + u_0^+ f_+ + \xi \sigma_3 u_3^+ f_- + \frac{\xi}{b} u_1^+ \vec{\sigma} \cdot [\vec{\ell}_z \times \nabla f_-], \quad (S1)$$

$$f_\pm = \sum_{m=1\ldots 6} (\pm 1)^{m+\frac{1}{2}} e^{i \vec{b}_m \cdot \vec{r}} \quad ,$$

where $\sigma_i$ are the Pauli matrices acting on the sublattice Bloch states $(\phi_{AK}, \phi_{BK})^T$ in the K valley ($\xi = 1$) and $(\phi_{BK'}, -\phi_{AK'})^T$ in the K' valley ($\xi = -1$). $f_\pm$ are the six shortest moiré Bragg vectors $\vec{b}_m$ ($b_{1,2,3,4,5,6} = b = \frac{4\pi}{\sqrt{3}a}$) of the superlattice, and $u_0$, $u_1$ $u_3$ are phenomenological parameters' which control the strength of the potential. In our calculations we used $u_0^+ = 21.7$ meV, $u_1^+ = -30.6$ meV and $u_3^+ = -22.2$ meV, which is based on previous works that predict the superlattice strength is related to the alignment angle and lattice mismatch of the graphene/hBN lattices (Ref. 7).

In the presence of magnetic field, the Dirac term is modified by the vector potential $\vec{A} = \frac{Bx_1}{a\sqrt{3}}(2\vec{a}_2 - \vec{a}_1)$ and incorporated into the momentum as $\vec{p} = -i\hbar\nabla + e\vec{A}$. We note that $\vec{A}$ is expressed in a hexagonal co-ordinate system $(x_1, x_2)$, such that $\vec{r} = x_1\vec{a}_1 + x_2\vec{a}_2$, to reflect the hexagonal symmetry of the moiré potential in graphene/hBN superlattices, with $\vec{a}_1, \vec{a}_2$ being the basis vectors of the Bravais lattice of the moiré pattern. Without any superlattice, the spectrum consists of infinitely degenerate Landau levels. Mathematically, this is determined by the fact that the group of translations in a magnetic field is non-Abelian[8], since the Aharonov-Bohm[9] effect introduces additional phase factors into the electronic wave function. In the case of the moiré superlattice, for magnetic field $B = \phi_0/S$ $(p/q)$, the Aharonov-Bohm phase attains quantized values of $2\pi$, and the electronic spectrum can then be described by Wannier states which propagate on a supercell that is $q$ times larger than the moiré unit cell (referred to as magnetic Bloch states in the main text). For the analysis of such states, we use the Landau levels to construct a basis set of Bloch-like states. This is done for each point in the miniature Brillouin zone of the corresponding supercell. We then diagonalize the Hamiltonian $\hat{H}$ numerically in that basis, checking the results for convergence against increasing the basis size. The details of such computations and examples of the resulting magnetic miniband spectra can be found in Refs. 7, 10 & 11.

**S4. Electron transport in the magnetic-minibands**

For $\phi/\phi_0 = p/q$, the electrons propagate as if they are in effectively zero magnetic field with a velocity that is determined by details of their magnetic miniband structures (Fig. 4b). This behavior causes local maxima in $\sigma_{xx}$ to appear at $\phi/\phi_0 = p/q$, with an amplitude that is governed by the Einstein conductivity formula (see Supplementary Information in Ref. 12)

$$\sigma_{xx} = \frac{4e^2}{h}\frac{E_F\tau}{\hbar}\frac{\langle v^2 \rangle}{v_F^2} \quad (S2)$$

where $\tau$ is the scattering time, $v_F$ is the Fermi-velocity of graphene, and $v$ is the group velocity of carriers at the Fermi-energy ($E_F$) in a particular magnetic Bloch state. We note that temperatures $T = 100 - 200$ K are not insignificant if compared with the width of magnetic-minibands (Fig. 4b), such that thermal smearing could populate carriers with a markedly different velocity to those at the Fermi-energy. Therefore, we consider *v* averaged over an interval of $\pm k_BT$ around the Fermi-energy, which enters equation (S2) as the mean-square velocity $\langle v^2 \rangle$. Assuming $\tau$ is constant[13], $\langle v^2 \rangle$ is the only variable that changes with magnetic field due to the varying miniband structures of magnetic Bloch states that form at different *B*. We expect the relative amplitudes of local maxima correlate directly with the changes in $\langle v^2 \rangle$ expected for different magnetic Bloch states. We extracted $\langle v^2 \rangle$ by calculating the group velocity for all *E* within $\pm k_BT$ of $E_F$ and taking the average. The group velocity at each *E* was determined by the familiar expression

$$\vec{v} = \frac{1}{\hbar} \frac{\partial E(k)}{\partial \vec{k}} \qquad (S3)$$

We calculated $\langle v^2 \rangle$ along a number of arbitrary directions and found the values differed only by about 1 %, which simply reflects a numerical error determined by the discrete set of points in the miniature Brillouin zones used in our simulations.

### S5. Hierarchy of magnetic Bloch states in the valence band

In the main text, we discussed the visibility of magnetic Bloch states in the conduction band, with regard to transport experiments. In general, we found that magnetic Bloch states with larger *p* had a smaller group velocity (Fig. 4c of main text). This results in smaller amplitude of $\sigma_{xx}$ (Fig. 2 of the main text). In experiment, the same qualitative trend was found for hole doping (Fig. 2d). For completeness, we calculated $\langle v^2 \rangle$ for magnetic Bloch states in the valence band at various $\phi/\phi_0 = p/q$ for a given *n* (Fig. S3). We found the same qualitative trend such that high-order states (larger *p* indicated by red and blue symbols in Fig. S3) have a systematically smaller $\langle v^2 \rangle$ as compared to the first-order states (black spheres). This demonstrates that the apparent hierarchy of states has a universal behavior for all kind of doping in the graphene/hBN spectrum.

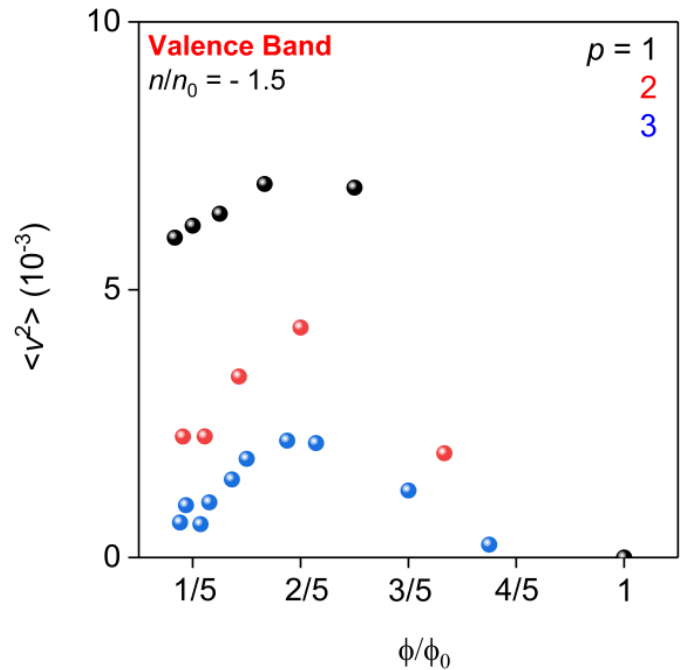

**Figure S3| Group velocity of holes in magnetic minibands.** The numerically calculated mean square velocity $<v^2>$ as a function of $\phi/\phi_0$ for different $p$. Here, we used moiré superlattice parameters listed in Section 3. The black, red and blue symbols correspond to magnetic states with $p$ = 1, 2 and 3 respectively.

**Supplementary References**

1. Mayorov, A. S. *et al.* Micrometer-Scale Ballistic Transport in Encapsulated Graphene at Room Temperature. *Nano Lett.* **11,** 2396–2399 (2011).

2. Kretinin, A. V *et al.* Electronic Properties of Graphene Encapsulated with Different Two-Dimensional Atomic Crystals. *Nano Lett.* **14,** 3270–3276 (2014).

3. Yankowitz, M. *et al.* Emergence of superlattice Dirac points in graphene on hexagonal boron nitride. *Nat Phys* **8,** 382–386 (2012).

4. Wang, L. *et al.* One-Dimensional Electrical Contact to a Two-Dimensional Material. *Science* **342,** 614 LP-617 (2013).

5. Ben Shalom, M. *et al.* Quantum oscillations of the critical current and high-field superconducting proximity in ballistic graphene. *Nat. Phys.* **12,** 318 (2015).

6. JuL. *et al.* Photoinduced doping in heterostructures of graphene and boron nitride. *Nat Nano* **9,** 348–352 (2014).